# Study of Sparsity-Aware Subband Adaptive Filtering Algorithms with Adjustable Penalties

Yi Yu[1], Haiquan Zhao[2], Rodrigo C. de Lamare[3], and Lu Lu[4]

*Abstract*—We propose two sparsity-aware normalized subband adaptive filter (NSAF) algorithms by using the gradient descent method to minimize a combination of the original NSAF cost function and the $l_1$-norm penalty function on the filter coefficients. This $l_1$-norm penalty exploits the sparsity of a system in the coefficients update formulation, thus improving the performance when identifying sparse systems. Compared with prior work, the proposed algorithms have lower computational complexity with comparable performance. We study and devise statistical models for these sparsity-aware NSAF algorithms in the mean square sense involving their transient and steady -state behaviors. This study relies on the vectorization argument and the paraunitary assumption imposed on the analysis filter banks, and thus does not restrict the input signal to being Gaussian or having another distribution. In addition, we propose to adjust adaptively the intensity parameter of the sparsity attraction term. Finally, simulation results in sparse system identification demonstrate the effectiveness of our theoretical results.

1. Introduction

Adaptive filtering is a branch of modern signal processing that has a variety of practical applications such as system identification, channel equalization, and acoustic/network echo cancellation [1]-[3]. The normalized least mean square (NLMS) is one of the most attractive algorithms, since it is robust to the input signal power and is very simple. Nonetheless, the problem of this algorithm is slow convergence when the input signal is highly correlated. To overcome this problem, various subband adaptive filters (SAFs) have been proposed in [4]-[6]. In SAFs, the input signal is divided into almost mutually exclusive multiple subband signals through the analysis filter, and then the decimated subband input signals (which are close to white signals) are used to update the adaptive filter coefficients, thus improving the convergence rate. It is worth noting that the multiband structure of SAF provides better performance in comparison with the conventional subband structure, since the former eliminates the aliasing and band edge effects [5], [6]. On the basis of this multiband structure, Lee and Gan presented the normalized SAF (NSAF) algorithm, which updates a fullband adaptive filter by using the decimated subband signals normalized by their respective subband input power [6]. In addition to its faster convergence rate for correlated input signals, the NSAF algorithm has comparable computational load to the NLMS algorithm especially for long filters. Hence, one of the interesting applications for the NSAF algorithm is acoustic echo cancelation. Following this algorithm, a large number of variants to further improve the performance of NSAF have been reported in the literature [7]-[10]. For example, to simultaneously achieve fast convergence rate and small final estimation error, several variable step size techniques have been developed from different optimization criteria [7], [8].

---
[1] School of Information Engineering, Southwest University of Science and Technology, Mianyang, China. (e-mail: yuyi_xyuan@163.com)
[2] School of Electrical Engineering, Southwest Jiaotong University, Chengdu, 610031, China. (hqzhao_swjtu@126.com).
[3] CETUC, PUC-Rio, Rio de Janeiro 22451-900, Brazil, and Department of Electronics, University of York, York YO10 5DD, U.K. (e-mail: rcdl500@ohm.york.ac.uk).
[4] School of Electronics and Information Engineering, Sichuan University, Chengdu, China. (lulu19900303@126.com).

In many practical scenarios, parameters of interest to be estimated are sparse, e.g., the echo paths [11] and digital TV transmission channels [12]. A property of sparse systems is that most of its entries have very small or zero magnitude. For such systems, traditional algorithms (e.g., the NLMS and the NSAF) suffer from slow convergence, since they do not take advantage of the sparsity of systems. To deal with this issue, one of the popular algorithms is the proportionate family, whose core idea is to assign an independent step size for each filter coefficient, which is proportional to the magnitude of that coefficient [3], [13]-[15], e.g., the proportionate NASF algorithm [15]. Alternatively, motivated by the compressive sensing framework [16], sparsity-aware algorithms has been proposed in [17]-[27]. To exploit the sparsity of the underlying system, sparsity-aware algorithms are obtained by adding a penalty function based on the $l_p$-norm of the filter coefficients to the objective function deriving the original algorithms, where $p=0$, 1, or $0<p<1$. This approach was firstly applied to the standard LMS, and examples of resulting algorithms are the zero-attraction LMS (ZA-LMS) based on the $l_1$-norm [17], the reweighted ZA-LMS (RZA-LMS) [17], and the $l_0$-LMS [18], [19], which exhibit better performance than the LMS algorithm for sparse systems. So far, the above sparsity-aware approaches have also been extended to other types of adaptive algorithms, e.g., the affine projection [20], [21] and distributed LMS [24] algorithms. Likewise, to improve the performance of NSAF algorithm in sparse cases, Choi developed its sparsity-aware versions, i.e., the $l_1$-norm based NSAF ($l_1$-NSAF) and reweighted $l_1$-NSAF ($l_1$-RNSAF) algorithms [23]. These two algorithms outperform the NSAF algorithm in terms of convergence rate and steady-state error when the system to be identified is sparse.

It is well-known that performance analysis of adaptive algorithms is always a research topic with many previously reported works in the literature, due to providing theoretical basis for designing adaptive filters [28]-[33]. The stochastic behavior of the NSAF algorithm was widely studied including the transient and steady-state performance [34]-[36]. To the best of our knowledge, however, there is no available performance analysis on the sparsity-aware NSAF algorithms. Therefore, the contributions of this paper are as follows:

1) For the $l_1$-NSAF and $l_1$-RNSAF algorithms, their low computational complexity versions are designed by directly relaxing the sparsity term in the update formulas, respectively, with almost similar convergence performance.

2) We analyze the asymptotic behavior of the proposed algorithms in the mean-square-deviation (MSD) sense, by using the vectorization argument presented in [1] that does not restrict the distribution of input signal. Similar analysis approaches have been used for the sparsity-aware LMS algorithms [29], [31], thus we refer to it for accomplishing the analyses of the proposed algorithms. Moreover, the MSD behaviors of the existing $l_1$-NSAF and $l_1$-RNSAF algorithms are also available based on the proposed analysis framework.

3) To further improve the performance of the proposed algorithms, an adaptive strategy for selecting the intensity parameter controlling the effect of the sparsity term is also proposed based on the MSD minimization.

Throughout the paper, some notations are adopted: $(\cdot)^T$ denotes the transpose of a matrix or vector; $\|\cdot\|_1$ and $\|\cdot\|_2$ denote the $l_1$- and $l_2$-norm of a vector, respectively; $E\{\cdot\}$ is to take the mathematical expectation; $\lambda_{\max}(\cdot)$ is the largest eigenvalue of a matrix; and $\mathrm{tr}(\cdot)$ is the trace of a matrix. The symbol $\mathbf{I}$ denotes the identity matrix of appropriate dimensions. In addition, all vectors are column vectors.

The remainder of this paper is organized as follows. In the next section, we briefly review the original sparsity-aware NSAF algorithms, and then propose their low complexity forms. In Section 3, the performance of the sparsity-aware NSAF algorithms described in the previous section is analyzed. Section 4 presents an adaptive intensity parameter strategy. In Section 5, simulations are performed to verify our works. Finally, conclusions are presented in Section 6.

2. Sparsity-aware NSAF algorithms

The desired signal $d(n)$ follows the model

$$d(n) = \mathbf{u}^T(n)\mathbf{w}^o + \eta(n), \tag{1}$$

where $\mathbf{w}^o = [w_1^o, w_2^o, ..., w_M^o]^T$ is an $M$-dimensional vector that we want to estimate, $\mathbf{u}(n) = [u(n), u(n-1), ..., u(n-M+1)]^T$ is the input vector, and $\eta(n)$ is the measurement noise. Fig. 1 shows the multiband structure of SAF with $N$ subbands. The input signal $u(n)$ and the desired signal $d(n)$ are partitioned into multiple subband signals $u_i(n)$ and $d_i(n)$ via the analysis filter bank $H_i(z)$, $i = 0,1,...,N-1$, respectively. The output signal $y_i(n)$ of the $i$th subband is obtained by filtering the signal $u_i(n)$ through an adaptive filter whose weight vector is denoted by $\mathbf{w}(k) = [w_1(k), w_2(k), ..., w_M(k)]^T$. Then, the signals $y_i(n)$ and $d_i(n)$ are $N$-fold decimated to yield signals $y_{i,D}(k)$ and $d_{i,D}(k)$ with a lower sampling rate, respectively. Here, we use $n$ to indicate the original sequences and $k$ to indicate the decimated sequences. Thus, the decimated subband error signals $e_{i,D}(k)$, $i = 0,1,...,N-1$ are expressed as

$$e_{i,D}(k) = d_{i,D}(k) - \mathbf{u}_i^T(k)\mathbf{w}(k) \tag{2}$$

where $\mathbf{u}_i(k) = [u_i(kN), u_i(kN-1), ..., u_i(kN-M+1)]^T$ and $d_{i,D}(k) = d_i(kN)$.

## 2.1 The original sparsity-aware NSAF algorithms

The original sparsity-aware NSAF update is derived by solving a minimum perturbation problem [23], as follows

$$\min_{\mathbf{w}(k+1)} \left\{ \|\mathbf{w}(k+1) - \mathbf{w}(k)\|_2^2 + \rho F(\mathbf{w}(k+1)) \right\} \tag{3}$$

subject to

$$d_{i,D}(k) - \mathbf{u}_i^T(k)\mathbf{w}(k+1) = 0 \text{ for } i = 0,1,...,N-1 \tag{4}$$

where $F(\cdot)$ indicates the penalty function based on the $l_1$-norm, and $\rho > 0$ controls the intensity given to the penalty function.

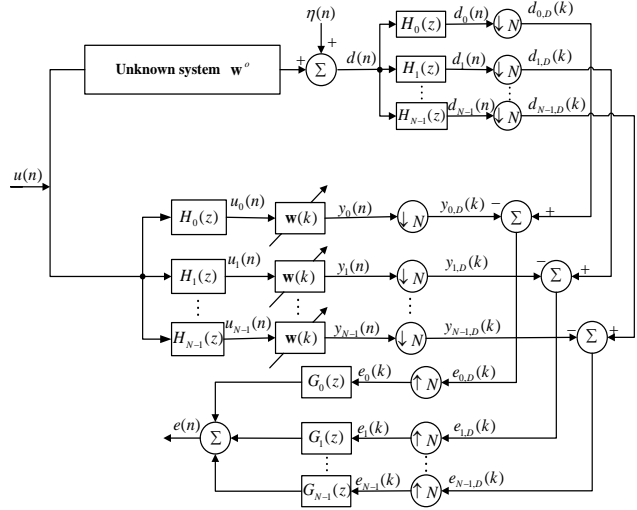

Fig. 1. Multiband-structure of SAF.

Applying the Lagrange multipliers method and a known *approximation* $\mathbf{u}_i^T(k)\mathbf{u}_j(k) \approx 0$, $i \neq j$ in the NSAF-type [6], the weight vector is updated as:

$$\mathbf{w}(k+1) = \mathbf{w}(k) + \mu \sum_{i=0}^{N-1} \frac{e_{i,D}(k)\mathbf{u}_i(k)}{\|\mathbf{u}_i(k)\|_2^2} + \frac{1}{2}\mu\rho \left( \sum_{i=0}^{N-1} \frac{\mathbf{u}_i(k)\mathbf{u}_i^T(k)}{\|\mathbf{u}_i(k)\|_2^2} - \mathbf{I}_M \right) f(\mathbf{w}(k)), \tag{5}$$

where $\mu$ is the step-size, and

$$\begin{aligned} f(\mathbf{w}(k)) &= \frac{\partial F(\mathbf{w}(k))}{\partial \mathbf{w}(k)} \\ &= \left[ \frac{\partial F(\mathbf{w}(k))}{\partial w_1(k)}, \frac{\partial F(\mathbf{w}(k))}{\partial w_2(k)}, ..., \frac{\partial F(\mathbf{w}(k))}{\partial w_M(k)} \right]^T \\ &\square \left[ f(w_1(k)), f(w_2(k)), ..., f(w_M(k)) \right]^T. \end{aligned} \tag{6}$$

In the derivation of (5), $\mathbf{w}(k+1)$ in $F(\mathbf{w}(k+1))$ is unavailable at the $k$th iteration, but it can be approximated by the current weight vector $\mathbf{w}(k)$, since it is expected that $\mathbf{w}(k+1)$ and $\mathbf{w}(k)$, near convergence, are in the same hyper-quadrant [25].

If the $l_1$-norm is considered for the penalty function, i.e., $F(\mathbf{w}(k)) = \|\mathbf{w}(k)\|_1 = \sum_{m=1}^{M} |w_m(k)|$, which leads to its gradient terms,

$$f(w_m(k)) = \text{sgn}(w_m(k)), m = 1, 2, ..., M,  \qquad (7)$$

then the $l_1$-NSAF algorithm is obtained, where $\text{sgn}(\cdot)$ denotes the sign function [23].

In the $l_1$-RNSAF algorithm, the reweighted gradient terms are expressed as [23]:

$$f(w_m(k)) = \frac{\text{sgn}(w_m(k))}{|w_m(k)| + \varepsilon}, m = 1, 2, ..., M, \qquad (8)$$

which can be obtained from the log-penalty function, i.e., $F(\mathbf{w}(k)) = \sum_{m=1}^{M} \ln(1 + |w(k)|/\varepsilon)$, where $\varepsilon > 0$ denotes the shrinkage magnitude.

**Remark 1:** In comparison with the NSAF algorithm, the sparsity-aware NSAF algorithms add the last term in (5), referred to as a zero-attractor (ZA), whose role is to pull small weight coefficients to zeros, thus improving the convergence rate of those coefficients. As shown in (7), the $l_1$-norm based ZA only considers the sign of the weight coefficients so that it shrinks uniformly all the coefficients to zeros without distinguishing zero and non-zero coefficients, thus it is not reasonable for large coefficients. However, the log-penalty function based ZA, i.e., (8), and $l_0$-norm based ZA can attract each small weight coefficient with an individual intensity being inversely proportional to its own magnitude, thus the resulting algorithms are superior to the $l_1$-norm based algorithm for sparse systems. Intuitively, although the $l_0$-norm can better characterize the sparsity of the underlying system than the log-penalty function, the $l_0$-norm is an NP-hard problem since it accounts for the number of nonzero coefficients. Thus, in practice, some continuous functions are considered to approximate the $l_0$-norm [21]. Roughly speaking, this log-penalty function is also a simple approximation of the $l_0$-norm [25], [37]. Without loss of generality, we can also incorporate the approximation function of the $l_0$-norm into the update (5) and/or the below update (9) to improve the NSAF' performance, but it is not the main purpose of this paper, since such an extension is straightforward only by substituting $f(\cdot)$. In addition, the performance of the $l_0$-norm based algorithms could also be analyzed along the analysis procedures described in Section 3, and the difference is how to deal with the moments involving the function $f(\cdot)$ in the analysis, which is left in our future work.

2.2 Proposed algorithms

As stated in Remark 1, the ZA term in (5) is to make use of the sparsity of the underlying system to obtain an improvement in performance. However, any $(m, j)$-th entry of the matrix $\sum_{i=0}^{N-1} \left( \mathbf{u}_i(k) \mathbf{u}_i^T(k) / \|\mathbf{u}_i(k)\|_2^2 \right)$ in this ZA term can be represented as $\sum_{i=0}^{N-1} \left( u_i(kN-m) u_i(kN-j) / \sum_{r=0}^{M-1} u_i^2(kN-r) \right)$, whose magnitude is less than and equal to $N/M$ so that it is much less than 1 especially for a long SAF [36], [38]. Furthermore, at each subband, the decimated subband input $u_i(kN)$ is close to the white signal for large $N$ [32]. It follows that, in contrast with $\mathbf{I}_M$, the matrix $\sum_{i=0}^{N-1}\left(\mathbf{u}_i(k)\mathbf{u}_i^T(k)\big/\|\mathbf{u}_i(k)\|_2^2\right)$ in (5) can be neglected. In doing so, (5) is simplified as:

$$\mathbf{w}(k+1) = \mathbf{w}(k) + \mu \sum_{i=0}^{N-1} \frac{e_{i,\mathrm{D}}(k)\mathbf{u}_i(k)}{\|\mathbf{u}_i(k)\|_2^2} - \frac{1}{2}\mu\rho f(\mathbf{w}(k)). \tag{9}$$

As a result, by combining (9) with (7) and (8), respectively, the proposed algorithms are developed, and correspondingly named as the quasi $l_1$-NSAF ($l_1$-qNSAF) and quasi $l_1$-RNSAF ($l_1$-qRNSAF) algorithms.

**Remark 2**: Table 1 provides the computational complexity of the algorithms for each fullband input sample, where the comparison operation is counted as the addition operation. Compared with the original $l_1$-NSAF and $l_1$-RNSAF algorithms, (9) based the proposed $l_1$-qNSAF and $l_1$-qRNSAF algorithms save the computational amount, respectively, without loss of performance as shown in Section 5.1. This amount stems from the calculation of $\left(\sum_{i=0}^{N-1}\mathbf{u}_i(k)\mathbf{u}_i^T(k)\big/\|\mathbf{u}_i(k)\|_2^2\right)f(\mathbf{w}(k))$ in (5), which requires $2M$ multiplications, $2M-M/N-1$ additions and 1 division. In addition, subband algorithms require additional $3NL$ multiplications and $3N(L-1)$ additions relative to the fullband adaptive algorithms (e.g., the NLMS), where $L$ denotes the length of the analysis filters.

Table 1 Computational complexity of the algorithms for each fullband input sample.

| Algorithms | Multiplications | Additions | Divisions |
|---|---|---|---|
| NLMS | $3M+1$ | $3M-1$ | 1 |
| NSAF | $3M+1+3NL$ | $3M-1+3N(L-1)$ | 1 |
| $l_1$-NSAF | $5M+1+3NL$ | $5M-2+3N(L-1)$ | 2 |
| $l_1$-RNSAF | $5M+1+3NL$ | $5M-2+1/N+3N(L-1)$ | $2+M/N$ |
| $l_1$-qNSAF | $3M+1+3NL$ | $3M-1+M/N+3N(L-1)$ | 1 |
| $l_1$-qRNSAF | $3M+1+3NL$ | $3M-1+(M+1)/N+3N(L-1)$ | $1+M/N$ |

It is worth noting that the update (9) can be also derived by minimizing the cost function $J(k)$ that is a combination of the NSAF's cost function [39] and the penalty function $F(\mathbf{w}(k))$, formulated as:

$$J(k) = \frac{1}{2}\sum_{i=0}^{N-1} \lambda_i e_{i,\mathrm{D}}^2(k) + \frac{1}{2}\rho F(\mathbf{w}(k)). \tag{10}$$

where $\lambda_i = 1/\|\mathbf{u}_i(k)\|_2^2$ denotes the weighting factor at the $i$-th subband. Then, applying the instantaneous gradient descent method, we can arrive at the update (9), given by:

$$\mathbf{w}(k+1) = \mathbf{w}(k) - \mu \frac{\partial J(k)}{\partial \mathbf{w}(k)}. \tag{11}$$

**Remark 3**: From another viewpoint, the ZA term in (5) is projected onto the subspace that is orthogonal to the space spanned by $\{\mathbf{u}_i(k), i = 0,1,...,N-1\}$ by pre-multiplying a $M \times M$ matrix which has a rank of $(M-N)$. Similar to [21], hence, such a projection manipulation restricts the influence of $f(\mathbf{w}(k))$ to a $(M-N)$ dimensional space, i.e., losing $N$ degrees of freedom, to guarantee the constraint equations (4) at each update when the step size equals 1. Since the proposed update (9) does not impose the constraints (4), it can be considered as a simple-relaxation implementation of (5) with being a more flexible update process.

**Remark 4**: According to (5) and (9), a unified update formula that describes the $l_1$-NSAF, $l_1$-RNSAF, $l_1$-qNSAF, and $l_1$-qRNSAF algorithms can be expressed as:

$$\mathbf{w}(k+1) = \mathbf{w}(k) + \mu \sum_{i=0}^{N-1} \frac{e_{i,\mathrm{D}}(k)\mathbf{u}_i(k)}{\|\mathbf{u}_i(k)\|_2^2} - \beta \mathbf{P}(k) f(\mathbf{w}(k)), \tag{12}$$

where $\mathbf{P}(k)$ and $f(\mathbf{w}(k))$ for these algorithms are given in Table 2, and $\beta = \mu\rho/2$ is referred to as the intensity parameter here.

Table 2  $\mathbf{P}(k)$ and $f(\mathbf{w}(k))$ for four sparsity-aware algorithms.

| Algorithms | $\mathbf{P}(k)$ | $f(w_m(k))$, $m=1,2,...,M$ |
|---|---|---|
| $l_1$-NSAF | $\mathbf{P}(k) = \mathbf{I}_M - \sum_{i=0}^{N-1} \frac{\mathbf{u}_i(k)\mathbf{u}_i^T(k)}{\|\mathbf{u}_i(k)\|_2^2}$ | $\mathrm{sgn}(w_m(k))$ |
| $l_1$-RNSAF |  | $\frac{\mathrm{sgn}(w_m(k))}{|w_m(k)|+\varepsilon}$ |
| $l_1$-qNSAF | $\mathbf{P}(k) = \mathbf{I}_M$ | $\mathrm{sgn}(w_m(k))$ |
| $l_1$-qRNSAF |  | $\frac{\mathrm{sgn}(w_m(k))}{|w_m(k)|+\varepsilon}$ |

3. Performance analyses of sparsity-aware algorithms

In this section, we will show how to carry out the performance analyses of the above-mentioned sparsity-aware NSAF algorithms in a unified way. For the analysis to be tractable, we rely on the following assumptions which are widely used in the performance analyses of SAF algorithms.

**Assumption 1**: The measurement noise $\eta(n)$ is a white process with zero-mean and variance $\sigma_\eta^2$.

**Assumption 2**: The filter banks for partitioning the input signal $u(n)$ and the desired signal $d(n)$ are assumed to be identical and paraunitary [32]-[36]. This assumption is to avoid the computation of the filter banks in the performance analyses. Thus, we can obtain the relation for subband index $i = 0,1,...,N-1$:

$$d_{i,D}(k) = \mathbf{u}_i^T(k)\mathbf{w}^o + \eta_i(k), \tag{13}$$

where $\eta_i(k)$ represents the *i*-th subband noise that can also be obtained from $\eta(n)$ through the analysis filter bank. Moreover, $\eta_i(k)$ is zero-mean white with variance $\sigma_{\eta_i}^2 = \sigma_\eta^2/N$.

**Assumption 3**: The subband noises $\eta_i(k)$, the past weight vectors $\mathbf{w}(l)$, $l \leq k$, and the decimated subband input vectors $\mathbf{u}_i(k)$ are statistically independent. This assumption is the well-known independence assumption [1], [28]-[36].

3.1 Transient behavior

Defining the weight error vector, $\tilde{\mathbf{w}}(k) = \mathbf{w}^o - \mathbf{w}(k)$, and then (12) is rewritten as

$$\tilde{\mathbf{w}}(k+1) = \tilde{\mathbf{w}}(k) - \mu \sum_{i=0}^{N-1} \frac{e_{i,\mathrm{D}}(k)\mathbf{u}_i(k)}{\|\mathbf{u}_i(k)\|_2^2} + \beta \mathbf{P}(k) f(\mathbf{w}(k)). \tag{14}$$

Substituting (2) and (13) into (14), we obtain

$$\tilde{\mathbf{w}}(k+1) = (\mathbf{I}_M - \mu \mathbf{A}(k))\tilde{\mathbf{w}}(k) - \mu \mathbf{B}(k) + \beta \mathbf{P}(k) f(\mathbf{w}(k)), \tag{15}$$

where

$$\mathbf{A}(k) = \sum_{i=0}^{N-1} \frac{\mathbf{u}_i(k)\mathbf{u}_i^T(k)}{\|\mathbf{u}_i(k)\|_2^2}, \tag{16}$$

and

$$\mathbf{B}(k) = \sum_{i=0}^{N-1} \frac{\eta_i(k)\mathbf{u}_i(k)}{\|\mathbf{u}_i(k)\|_2^2}. \tag{17}$$

Using the assumption 3 and taking the expectation of (15), we find the mean evolution formulation of $\tilde{\mathbf{w}}(k)$ as

$$E\{\tilde{\mathbf{w}}(k+1)\} = (\mathbf{I}_M - \mu E\{\mathbf{A}(k)\}) E\{\tilde{\mathbf{w}}(k)\} + \beta E\{\mathbf{P}(k)\} E\{f(\mathbf{w}(k))\}, \tag{18}$$

where $E\{\mathbf{P}(k)\}$ is given in Table 3.

Multiplying (15) by its transpose, we obtain:

$$\begin{aligned}
\tilde{\mathbf{w}}(k+1)\tilde{\mathbf{w}}^T(k+1) = &\ (\mathbf{I}_M - \mu \mathbf{A}(k))\tilde{\mathbf{w}}(k)\tilde{\mathbf{w}}^T(k)(\mathbf{I}_M - \mu \mathbf{A}(k))^T \\
&- \mu (\mathbf{I}_M - \mu \mathbf{A}(k))\tilde{\mathbf{w}}(k)\mathbf{B}^T(k) \\
&+ \beta (\mathbf{I}_M - \mu \mathbf{A}(k))\tilde{\mathbf{w}}(k) f^T(\mathbf{w}(k)) \mathbf{P}^T(k) \\
&- \mu \mathbf{B}(k)\tilde{\mathbf{w}}^T(k)(\mathbf{I}_M - \mu \mathbf{A}(k))^T \\
&+ \mu^2 \mathbf{B}(k)\mathbf{B}^T(k) - \mu\beta \mathbf{B}(k) f^T(\mathbf{w}(k)) \mathbf{P}^T(k) \\
&+ \beta \mathbf{P}(k) f(\mathbf{w}(k)) \tilde{\mathbf{w}}^T(k)(\mathbf{I}_M - \mu \mathbf{A}(k))^T \\
&- \mu\beta \mathbf{P}(k) f(\mathbf{w}(k)) \mathbf{B}^T(k) \\
&+ \beta^2 \mathbf{P}(k) f(\mathbf{w}(k)) f^T(\mathbf{w}(k)) \mathbf{P}^T(k).
\end{aligned} \tag{19}$$

Introducing the definition of the covariance matrix of the weight error vector: $\mathbf{\Phi}(k) \triangleq E\{\tilde{\mathbf{w}}(k)\tilde{\mathbf{w}}^T(k)\}$, and taking the expectation of both sides of (19) under assumptions 2 and 3, we can arrive at

$$\mathbf{\Phi}(k+1) = \underbrace{E\left\{\left(\mathbf{I}_M - \mu\mathbf{A}(k)\right)\mathbf{\Phi}(k)\left(\mathbf{I}_M - \mu\mathbf{A}(k)\right)^T\right\}}_{\text{I}}$$
$$+\beta \underbrace{E\left\{\left(\mathbf{I}_M - \mu\mathbf{A}(k)\right)\mathbf{H}(k)\mathbf{P}^T(k)\right\}}_{\text{II}}$$
$$+\beta \underbrace{E\left\{\mathbf{P}(k)\mathbf{H}^T(k)\left(\mathbf{I}_M - \mu\mathbf{A}(k)\right)^T\right\}}_{\text{III}}$$
$$+\beta^2 \underbrace{E\left\{\mathbf{P}(k)\mathbf{\Theta}(k)\mathbf{P}^T(k)\right\}}_{\text{IV}} + \mu^2 \mathbf{D}. \qquad (20)$$

where

$$\mathbf{H}(k) = E\{\tilde{\mathbf{w}}(k)f^T(\mathbf{w}(k))\}, \qquad (21)$$

$$\mathbf{\Theta}(k) = E\{f(\mathbf{w}(k))f^T(\mathbf{w}(k))\}, \qquad (22)$$

and

$$\mathbf{D} = E\{\mathbf{B}(k)\mathbf{B}^T(k)\}. \qquad (23)$$

Exploiting the assumption that different subband noises are mutually independent, i.e., $E\{\eta_i(k)\eta_j(k)\} = E\{\eta_i(k)\}E\{\eta_j(k)\}$, $i \neq j$, then (23) is simplified to

$$\mathbf{D} = \sum_{i=0}^{N-1} \sigma_{\eta_i}^2 E\left\{\frac{\mathbf{u}_i(k)\mathbf{u}_i^T(k)}{\|\mathbf{u}_i(k)\|_2^4}\right\} = \frac{\sigma_\eta^2}{N}\sum_{i=0}^{N-1} E\left\{\frac{\mathbf{u}_i(k)\mathbf{u}_i^T(k)}{\|\mathbf{u}_i(k)\|_2^4}\right\}. \qquad (24)$$

In order to further simplify (20), we will use the vectorization operation and the property of the Kronecker product [40]. So, the notation vec(·) means that a $M \times M$ matrix is converted to an $M^2 \times 1$ column vector whose entries are formed by stacking the successive columns of the matrix on top of each other, and $\text{vec}^{-1}(\cdot)$ is the inverse operation of vec(·). For any matrices $\{\mathbf{X}, \mathbf{\Sigma}, \mathbf{Y}\}$ of compatible dimensions, we have the property

$$\text{vec}(\mathbf{X}\mathbf{\Sigma}\mathbf{Y}) = (\mathbf{Y}^T \otimes \mathbf{X})\text{vec}(\mathbf{\Sigma}), \qquad (25)$$

where $\otimes$ denotes the Kronecker product. Based on the relation (25), we are able to reformulate the terms I-IV in (20), as shown in Appendix A. Then, applying (A1)-(A4), (20) can be expressed in a vector form:

$$\text{vec}(\mathbf{\Phi}(k+1)) = \mathbf{F}_1 \text{vec}(\mathbf{\Phi}(k)) + \beta \mathbf{F}_2 \text{vec}(\mathbf{H}(k)) \\ + \beta \mathbf{F}_3 \text{vec}(\mathbf{H}^T(k)) + \beta^2 \mathbf{F}_4 \text{vec}(\mathbf{\Theta}(k)) \\ + \mu^2 \text{vec}(\mathbf{D}). \tag{26}$$

where $\mathbf{F}_1$ is an $M^2 \times M^2$ matrix given by

$$\mathbf{F}_1 = \mathbf{I}_{M^2} - \mu(E\{\mathbf{A}(k)\} \otimes \mathbf{I}_M) - \\ \mu(\mathbf{I}_M \otimes E\{\mathbf{A}(k)\}) + \mu^2 E\{\mathbf{A}(k) \otimes \mathbf{A}(k)\}, \tag{27}$$

and $\mathbf{F}_2$, $\mathbf{F}_3$, and $\mathbf{F}_4$ are given in Table 3.

Table 3 Calculation of $E\{\mathbf{P}(k)\}$, $\mathbf{F}_2$, $\mathbf{F}_3$, and $\mathbf{F}_4$ for the sparsity-aware NSAF algorithms

|  | $l_1$-NSAF and $l_1$-RNSAF | $l_1$-qNSAF and $l_1$-qRNSAF |
|---|---|---|
| $E\{\mathbf{P}(k)\}$ | $\mathbf{I}_M - E\{\mathbf{A}(k)\}$ | $\mathbf{I}_M$ |
| $\mathbf{F}_2$ | $\mathbf{I}_{M^2} - (E\{\mathbf{A}(k)\} \otimes \mathbf{I}_M) - \mu(\mathbf{I}_M \otimes E\{\mathbf{A}(k)\}) + \mu E\{\mathbf{A}(k) \otimes \mathbf{A}(k)\}$ | $\mathbf{I}_{M^2} - \mu(\mathbf{I}_M \otimes E\{\mathbf{A}(k)\})$ |
| $\mathbf{F}_3$ | $\mathbf{I}_{M^2} - (\mathbf{I}_M \otimes E\{\mathbf{A}(k)\}) - \mu(E\{\mathbf{A}(k)\} \otimes \mathbf{I}_M) + \mu E\{\mathbf{A}(k) \otimes \mathbf{A}(k)\}$ | $\mathbf{I}_{M^2} - \mu(E\{\mathbf{A}(k)\} \otimes \mathbf{I}_M)$ |
| $\mathbf{F}_4$ | $\mathbf{I}_{M^2} - (E\{\mathbf{A}(k)\} \otimes \mathbf{I}_M) - (\mathbf{I}_M \otimes E\{\mathbf{A}(k)\}) + E\{\mathbf{A}(k) \otimes \mathbf{A}(k)\}$ | $\mathbf{I}_{M^2}$ |

The MSD at iteration $k$ is defined as:

$$\text{MSD}(k) \triangleq E\{\tilde{\mathbf{w}}^T(k)\tilde{\mathbf{w}}(k)\} = \text{tr}(\mathbf{\Phi}(k)). \tag{28}$$

As a consequence, the transient MSD behavior of the sparsity-aware NSAF algorithms can be described by (28) with the recursions (18) and (26). However, it still requires the computation of the moments $E\{f(\mathbf{w}(k))\}$, $\mathbf{H}(k)$ and $\mathbf{\Theta}(k)$ in (18) and (26) in advance. To accomplish this goal, we employ again two assumptions.

**Assumption 4**: The $m$-th entry of the weight error vector $\tilde{\mathbf{w}}(k)$ at iteration $k$, denoted by $\tilde{w}_m(k)$, has a Gaussian distribution with mean $z_m(k)$ and variance $\sigma_m^2(k)$, namely, $\tilde{w}_m(k) : \mathcal{N}(z_m(k), \sigma_m^2(k))$. This assumption has been used in the literature [19], [29], [30], and will be verified in Appendix B. According to this assumption, the distribution of the $m$-th weight coefficient is expressed as $w_m(k) = w_m^o - \tilde{w}_m(k) : \mathcal{N}(w_m^o - z_m(k), \sigma_m^2(k))$.

**Assumption 5**: When $m \neq j$, we can use the approximations $E\{f(w_m(k))f(w_j(k))\} \approx E\{f(w_m(k))\}E\{f(w_j(k))\}$ and $E\{\tilde{w}_m(k)f(w_j(k))\} \approx E\{\tilde{w}_m(k)\}E\{f(w_j(k))\}$. This is a separable assumption, which has been reported in [29], [30], [31].

**Case 1**: $f(w_m(k)) = \text{sgn}(w_m(k))$. In this case, using the assumption 5, $\mathbf{H}(k)$ and $\mathbf{\Theta}(k)$ can be rewritten in component form as: respectively,

$$\mathbf{H}_{m,j}(k) = \begin{cases} w_m^o E\{\text{sgn}(w_m(k))\} - E\{|w_m(k)|\}, & m = j \\ w_m^o E\{\text{sgn}(w_j(k))\} - E\{w_m(k)\} E\{\text{sgn}(w_j(k))\}, & m \neq j, \end{cases} \quad (29)$$

$$\mathbf{\Theta}_{m,j}(k) = \begin{cases} 1, & m = j \\ E\{\text{sgn}(w_m(k))\} E\{\text{sgn}(w_j(k))\}, & m \neq j. \end{cases} \quad (30)$$

where $(\cdot)_{m,j}$ denotes the $(m, j)$-th entry of a matrix.

Using the assumption 4, we can take the expectations in (29) and (30) [1]:

$$E\{w_m(k)\} = w_m^o - E\{\tilde{w}_m(k)\}, \quad (31)$$

$$E\{|w_m(k)|\} = (w_m^o - z_m(k))\text{erf}\left(\frac{w_m^o - z_m(k)}{\sqrt{2\sigma_m^2(k)}}\right) + \sqrt{\frac{2}{\pi}}\sigma_m(k)\exp\left(-\frac{(w_m^o - z_m(k))^2}{2\sigma_m^2(k)}\right), \quad (32)$$

$$E\{\text{sgn}(w_m(k))\} = -\text{erf}\left(-\frac{w_m^o - z_m(k)}{\sqrt{2\sigma_m^2(k)}}\right), \quad (33)$$

where the function erf($\cdot$) is defined as [30]:

$$\text{erf}(x) = \frac{2}{\sqrt{\pi}} \int_0^x \exp(-t^2) dt, \quad (34)$$

and the mean $z_m(k)$ and variance $\sigma_m^2(k)$ are calculated by, respectively,

$$z_m(k) = E\{\tilde{w}_m(k)\}, \quad (35)$$

and

$$\sigma_m^2(k) = \mathbf{\Phi}_{m,m}(k) - E^2\{\tilde{w}_m(k)\}. \quad (36)$$

**Case 2**: $f(w_m(k)) = \text{sgn}(w_m(k))/(|w_m(k)| + \varepsilon)$. Likewise, using the assumption 4, $\mathbf{H}(k)$ and $\mathbf{\Theta}(k)$ can be calculated respectively as

---

[1] $E\{|w_m(k)|\} = \frac{1}{\sqrt{2\pi}\sigma_m(k)} \int_{-\infty}^{\infty} |w_m(k)| \times \exp\left(-[w_m(k) - (w_m^o - z_m(k))]^2 / 2\sigma_m^2(k)\right) dw_m(k)$, and $E\{\text{sgn}(w_m(k))\} = \frac{1}{\sqrt{2\pi}\sigma_m(k)} \int_{-\infty}^{\infty} |w_m(k)| \times \exp\left(-[w_m(k) - (w_m^o - z_m(k))]^2 / 2\sigma_m^2(k)\right) dw_m(k)$.

$$\mathbf{H}_{m,j}(k) = \begin{cases} w_m^o E\left\{\dfrac{\text{sgn}(w_m(k))}{|w_m(k)|+\varepsilon}\right\} - E\left\{\dfrac{|w_m(k)|}{|w_m(k)|+\varepsilon}\right\}, & m = j \\ w_m^o E\left\{\dfrac{\text{sgn}(w_m(k))}{|w_m(k)|+\varepsilon}\right\} - E\{w_m(k)\} E\left\{\dfrac{\text{sgn}(w_m(k))}{|w_m(k)|+\varepsilon}\right\}, & m \neq j, \end{cases} \tag{37}$$

$$\mathbf{\Theta}_{m,j}(k) = \begin{cases} E\left\{\dfrac{1}{(|w_m(k)|+\varepsilon)^2}\right\}, & m = j \\ E\left\{\dfrac{\text{sgn}(w_m(k))}{|w_m(k)|+\varepsilon}\right\} E\left\{\dfrac{\text{sgn}(w_m(k))}{|w_m(k)|+\varepsilon}\right\}, & m \neq j. \end{cases} \tag{38}$$

Considering that the value of $\varepsilon$ is small, the expectations in (37) and (38) can be approximated as:

$$E\left\{\dfrac{|w_m(k)|}{|w_m(k)|+\varepsilon}\right\} \approx \dfrac{E\{|w_m(k)|\}}{E\{|w_m(k)|\}+\varepsilon}, \tag{39}$$

$$E\left\{\dfrac{\text{sgn}(w_m(k))}{|w_m(k)|+\varepsilon}\right\} \approx \dfrac{E\{\text{sgn}(w_m(k))\}}{E\{|w_m(k)|\}+\varepsilon}, \tag{40}$$

$$E\left\{\dfrac{1}{(|w_m(k)|+\varepsilon)^2}\right\} \approx \dfrac{1}{E\{w_m^2(k)\}+2\varepsilon E\{|w_m(k)|\}+\varepsilon^2}. \tag{41}$$

At this point, we have completed the transient analysis.

3.2 Stable convergence conditions

Based on the definitions in (7) and (8), one can find that $f(\mathbf{w}(k))$ has bounded entries [19], [24]. In particular, we have $\|f(\mathbf{w}(k))\|_2 \leq \sqrt{M}$ for the definition (7), and $\|f(\mathbf{w}(k))\|_2 \leq \sqrt{M}/\varepsilon$ for the definition (8). Also, the value of $\beta$ is usually small, as can be seen in the simulation section. Hence, to ensure the mean stability of the sparsity-aware NSAF algorithms, the maximum eigenvalue of the matrix $(\mathbf{I}_M - \mu E\{\mathbf{A}(k)\})$ in (18) must be less than 1, which leads to

$$0 < \mu < \dfrac{2}{\lambda_{\max}(E\{\mathbf{A}(k)\})}. \tag{42}$$

Since the matrices $\mathbf{H}(k)$ and $\mathbf{\Theta}(k)$ are also bounded (see [19], [24] for details), we can use *theorem 2* in [28] to obtain the mean square convergence condition of the sparsity-aware NSAF algorithms from (26). Specifically, the matrix $\mathbf{F}_1$ given by (27) must be stable, namely, $\lambda_{\max}(\mathbf{F}_1) < 1$, thereby acquiring the range of the step size $\mu$:

$$0 < \mu < \min\left\{\dfrac{2}{\lambda_{\max}(\mathbf{L}^{-1}\mathbf{\Psi})}, \dfrac{2}{\max(\lambda(\mathbf{\Xi}) \in R^+)}\right\}, \tag{43}$$

where $\mathbf{L} = E\{\mathbf{A}(k)\} \otimes \mathbf{I}_M + \mathbf{I}_M \otimes E\{\mathbf{A}(k)\}$, $\mathbf{\Psi} = E\{\mathbf{A}(k) \otimes \mathbf{A}(k)\}$, and $\mathbf{\Xi} = \begin{bmatrix} \mathbf{L}/2 & -\mathbf{\Psi}/2 \\ \mathbf{I} & \mathbf{0} \end{bmatrix}$.

Obviously, the stable convergence conditions of the sparsity-aware NSAF algorithms, i.e., (42) and (43), are the same as that of the NSAF algorithm presented in [33].

### 3.3 Steady-state behavior

In the steady-state, i.e., $k \to \infty$, from (18) we obtain:

$$E\{\tilde{\mathbf{w}}(\infty)\} = \frac{\beta}{\mu} \left(E\{\mathbf{A}(\infty)\}\right)^{-1} E\{\mathbf{P}(\infty)\} E\{f(\mathbf{w}(\infty))\}. \tag{44}$$

Thus, the mean weight vector $E\{\mathbf{w}(k)\}$ for the sparsity-aware NSAF algorithms converges to

$$E\{\mathbf{w}(\infty)\} = \mathbf{w}^o - \frac{\beta}{\mu} \left(E\{\mathbf{A}(\infty)\}\right)^{-1} E\{\mathbf{P}(\infty)\} E\{f(\mathbf{w}(\infty))\}. \tag{45}$$

According to the definition of $\mathbf{\Phi}(k)$, at the steady-state stage of the algorithms, we have $\mathbf{\Phi}(k+1) = \mathbf{\Phi}(k)$, $k \to \infty$. Therefore, the steady-state solution $\mathbf{\Phi}(\infty)$ of (26) is obtained:

$$\mathbf{\Phi}(\infty) = \text{vec}^{-1}\left(\left(\mathbf{I}_{M^2} - \mathbf{F}_1\right)^{-1} \begin{bmatrix} \beta \mathbf{F}_2 \text{vec}(\mathbf{H}(k)) + \beta \mathbf{F}_3 \text{vec}(\mathbf{H}^T(k)) \\ + \beta^2 \mathbf{F}_4 \text{vec}(\mathbf{\Theta}(k)) + \mu^2 \text{vec}(\mathbf{D}) \end{bmatrix}\right). \tag{46}$$

By resorting to the property $\text{tr}(\mathbf{XY}) = \left(\text{vec}(\mathbf{X}^T)\right)^T \text{vec}(\mathbf{Y})$, (46) is changed to

$$\begin{aligned} \text{MSD}(\infty) &\triangleq \text{tr}(\mathbf{\Phi}(\infty)) \\ &= \mu^2 \text{vec}^T(\mathbf{I}_M)\left(\mathbf{I}_{M^2} - \mathbf{F}_1\right)^{-1} \text{vec}(\mathbf{D}) + \Delta, \end{aligned} \tag{47}$$

where

$$\Delta = \text{vec}^T(\mathbf{I}_M)\left(\mathbf{I}_{M^2} - \mathbf{F}_1\right)^{-1} \begin{bmatrix} \beta \mathbf{F}_2 \text{vec}(\mathbf{H}(\infty)) + \beta \mathbf{F}_3 \text{vec}(\mathbf{H}^T(\infty)) \\ + \beta^2 \mathbf{F}_4 \text{vec}(\mathbf{\Theta}(\infty)) \end{bmatrix}. \tag{48}$$

Note that, if $\beta = 0$ (i.e., $\Delta = 0$), then (47) represents the steady-state MSD of the NSAF algorithm [33]. The sparsity-aware NSAF algorithms will outperform the NSAF algorithm when $\Delta < 0$; equivalently, the intensity parameter satisfy the following condition:

$$0 < \beta < \beta^* = \frac{-\text{vec}^T(\mathbf{I}_M)\left(\mathbf{I}_{M^2} - \mathbf{F}_1\right)^{-1}\left[\mathbf{F}_2 \text{vec}(\mathbf{H}(\infty)) + \mathbf{F}_3 \text{vec}(\mathbf{H}^T(\infty))\right]}{\text{vec}^T(\mathbf{I}_M)\left(\mathbf{I}_{M^2} - \mathbf{F}_1\right)^{-1} \mathbf{F}_4 \text{vec}(\mathbf{\Theta}(\infty))}. \tag{49}$$

The relation (49) reveals that the sparsity-aware NSAF algorithms have better steady-state performance only when the value of $\beta$ is in a particular range $(0, \beta^*)$. Moreover, this fact can be roughly observed in Section 5. 2.

Since (47) is a nonlinear equation, it is difficult to find a closed-form solution on MSD($\infty$). Hence, we only consider its numerical solution by using the following ways: 1) obtaining MSD($\infty$) from (26) in a recursive way; and 2) obtaining the approximate solution of MSD($\infty$) under some additional assumptions, as described in the sequel.

Rewriting (28) in a component-wise, we have

$$\text{MSD}(\infty) = \sum_{m=1}^{M} E\left\{\tilde{w}_m^2(\infty)\right\} = \sum_{m=1}^{M} \mathbf{\Phi}_{m,m}(\infty), \tag{50}$$

where from (46), $\mathbf{\Phi}_{m,m}(\infty)$ can be expressed as:

$$\begin{aligned}
\mathbf{\Phi}_{m,m}(\infty) = &\mu^2 \left[\text{vec}^{-1}\left(\left(\mathbf{I}_{M^2} - \mathbf{F}_1\right)^{-1} \text{vec}(\mathbf{D})\right)\right]_{m,m} + \\
&\beta \left[\text{vec}^{-1}\left(\left(\mathbf{I}_{M^2} - \mathbf{F}_1\right)^{-1} \mathbf{F}_2 \text{vec}(\mathbf{H}(\infty))\right)\right]_{m,m} + \\
&\beta \left[\text{vec}^{-1}\left(\left(\mathbf{I}_{M^2} - \mathbf{F}_1\right)^{-1} \mathbf{F}_3 \text{vec}(\mathbf{H}^T(\infty))\right)\right]_{m,m} + \\
&\beta^2 \left[\text{vec}^{-1}\left(\left(\mathbf{I}_{M^2} - \mathbf{F}_1\right)^{-1} \mathbf{F}_4 \text{vec}(\mathbf{\Theta}(\infty))\right)\right]_{m,m}.
\end{aligned} \tag{51}$$

**Assumption 6**: The decimated input signal of each subband can be assumed as a white signal if $N$ is large enough [32], [36] so that the matrices $\mathbf{F}_1$, $\mathbf{F}_2$, $\mathbf{F}_3$, $\mathbf{F}_4$, and $\mathbf{D}$ are diagonal. With this assumption, (51) is simplified as:

$$\begin{aligned}
\mathbf{\Phi}_{m,m}(\infty) = &\mu^2 \left[\text{vec}^{-1}\left(\left(\mathbf{I}_{M^2} - \mathbf{F}_1\right)^{-1} \text{vec}(\mathbf{D})\right)\right]_{m,m} + \\
&\beta \left(\left(\mathbf{I}_{M^2} - \mathbf{F}_1\right)^{-1}\left(\mathbf{F}_2 + \mathbf{F}_3\right)\right)_{1+(m-1)(M+1), 1+(m-1)(M+1)} \mathbf{H}_{m,m}(\infty) + \\
&\beta^2 \left(\left(\mathbf{I}_{M^2} - \mathbf{F}_1\right)^{-1} \mathbf{F}_4\right)_{1+(m-1)(M+1), 1+(m-1)(M+1)} \mathbf{\Theta}_{m,m}(\infty).
\end{aligned} \tag{52}$$

To proceed, we will classify the entries of the sparse vector $\mathbf{w}^o$ into two categories: the sets of the zero and nonzero entries, denoted by Z and NZ, respectively. Namely, $w_m^o = 0$ for $m \in \text{Z}$, and $w_m^o \neq 0$ for $m \in \text{NZ}$. Assuming that the step size is small, we have $|\tilde{w}_m| > w_m^o$ for $m \in \text{Z}$ and $|\tilde{w}_m| \ll w_m^o$ for $m \in \text{NZ}$ in the steady-state [31]. With this in mind, the following approximations can hold:

**Case 1:** if $f(w_m(k)) = \text{sgn}(w_m(k))$, then we have

$$\mathbf{H}_{m,m}(\infty) = \begin{cases} -E\left\{|\tilde{w}_m(\infty)|\right\}, & m \in \text{Z} \\ E\left\{\tilde{w}_m(\infty)\right\} \text{sgn}(w_m^o), & m \in \text{NZ}, \end{cases} \tag{53}$$

$$\mathbf{\Theta}_{m,m}(\infty) = 1, \tag{54}$$

$$E\{\tilde{w}_m(\infty)\} = \begin{cases} 0, & m \in Z \\ \dfrac{\beta}{\mu}\left(\left(E\{\mathbf{A}(k)\}\right)^{-1} E\{\mathbf{P}(k)\}\right)_{m,m} \mathrm{sgn}(w_m^o), & m \in \mathrm{NZ}. \end{cases} \tag{55}$$

**Case 2:** if $f(w_m(k)) = \mathrm{sgn}(w_m(k))/(|w_m(k)|+\varepsilon)$, then we get

$$\mathbf{H}_{m,m}(\infty) = \begin{cases} -E\left\{\dfrac{|\tilde{w}_m(\infty)|}{|\tilde{w}_m(\infty)|+\varepsilon}\right\}, & m \in Z \\ E\{\tilde{w}_m(\infty)\}\dfrac{\mathrm{sgn}(w_m^o)}{|w_m^o|+\varepsilon}, & m \in \mathrm{NZ}, \end{cases} \tag{56}$$

$$\mathbf{\Theta}_{m,m}(\infty) = \begin{cases} E\left\{\dfrac{1}{(|\tilde{w}_m(\infty)|+\varepsilon)^2}\right\}, & m \in Z \\ \dfrac{1}{(|w_m^o|+\varepsilon)^2}, & m \in \mathrm{NZ}, \end{cases} \tag{57}$$

$$E\{\tilde{w}_m(\infty)\} = \begin{cases} 0, & m \in Z \\ \dfrac{\beta}{\mu}\left(\left(E\{\mathbf{A}(k)\}\right)^{-1} E\{\mathbf{P}(k)\}\right)_{m,m}\dfrac{\mathrm{sgn}(w_m^o)}{|w_m^o|+\varepsilon}, & m \in \mathrm{NZ}. \end{cases} \tag{58}$$

Using Price's theorem [31], we get

$$E\{|\tilde{w}_m(\infty)|\} = \sqrt{2/\pi}\,\sigma_m(\infty) = \sqrt{2/\pi}\,\mathbf{\Phi}_{m,m}(\infty), \quad m \in Z. \tag{59}$$

Let $Q$ denote the number of the non-zero components in the sparse vector $\mathbf{w}^o$, i.e., the cardinality of NZ. Obviously, the smaller the value of $Q$ is, the sparser the vector $\mathbf{w}^o$ is. In the steady-state, it can be assumed that the variances of the weight error coefficients $\tilde{w}_m(k)$ for $m \in Z$ are identical, denoted as $\sigma_z^2 = \sigma_m^2(\infty), m \in Z$. Then, $\sigma_z^2$ can be obtained by solving the positive root of the equation

$$\begin{aligned}(M-Q)\sigma_z^2 &= \mu^2 \sum_{m \in z}\left[\mathrm{vec}^{-1}\left(\left(\mathbf{I}_{M^2} - \mathbf{F}_1\right)^{-1}\mathrm{vec}(\mathbf{D})\right)\right]_{m,m} + \\ &\quad \beta\sum_{m \in z}\left(\left(\mathbf{I}_{M^2} - \mathbf{F}_1\right)^{-1}(\mathbf{F}_2 + \mathbf{F}_3)\right)_{1+(m-1)(M+1),1+(m-1)(M+1)}\mathbf{H}_{m,m}(\infty) + \\ &\quad \beta^2 \sum_{m \in z}\left(\left(\mathbf{I}_{M^2} - \mathbf{F}_1\right)^{-1}\mathbf{F}_4\right)_{1+(m-1)(M+1),1+(m-1)(M+1)}\mathbf{\Theta}_{m,m}(\infty). \end{aligned} \tag{60}$$

In addition, the variances $\sigma_m^2(\infty)$ of the weight error coefficients for $m \in \mathrm{NZ}$ are obtained by solving the equation

$$\begin{aligned}\sigma_m^2(\infty) = \mu^2 &\left[\text{vec}^{-1}\left(\left(\mathbf{I}_{M^2} - \mathbf{F}_1\right)^{-1}\text{vec}(\mathbf{D})\right)\right]_{m,m} + \\ &\beta\left(\left(\mathbf{I}_{M^2} - \mathbf{F}_1\right)^{-1}\left(\mathbf{F}_2 + \mathbf{F}_3\right)\right)_{1+(m-1)(M+1),1+(m-1)(M+1)} \mathbf{H}_{m,m}(\infty) + \\ &\beta^2\left(\left(\mathbf{I}_{M^2} - \mathbf{F}_1\right)^{-1}\mathbf{F}_4\right)_{1+(m-1)(M+1),1+(m-1)(M+1)} \mathbf{\Theta}_{m,m}(\infty).\end{aligned} \tag{61}$$

Combining (50), (60) and (61), we formulate the steady-state MSD of the sparsity-aware NSAF algorithms as:

$$\text{MSD}(\infty) = (M-Q)\sigma_z^2 + \sum_{m \in \text{NZ}} \sigma_m^2(\infty). \tag{62}$$

4. Adaptation of the intensity parameter

As stated in Section 3.3, the intensity parameter $\beta$ must be chosen in the interval $(0, \beta^*)$ so that the sparsity-aware NSAF algorithms outperform the NSAF algorithm. Importantly, there is an optimal value $\beta_{\text{opt}}$ in this interval. However, choosing $\beta_{\text{opt}}$ is very difficult as it is relative to the unknown vector. To address similar $\beta$ problem in the fullband sparsity-aware algorithms, several adaptive techniques have been presented, e.g., [41]-[44]. However, as far as we know, it has not been studied in the subband algorithms. So, we propose to replace $\beta$ with the adaptation $\beta(k)$. Taking the squared $l_2$-norm of both sides of (15) with $\beta(k)$ and then enforcing its conditional expectation over $\mathbf{w}(k)$ under assumptions 2 and 3, we arrive at the following formula:

$$\begin{aligned}E\left\{\|\tilde{\mathbf{w}}(k+1)\|_2^2 \,\big|\, \mathbf{w}(k)\right\} = &\mu^2 E\left\{\mathbf{B}^T(k)\mathbf{B}(k)\right\} + \\ &\tilde{\mathbf{w}}^T(k)E\left\{\left(\mathbf{I}_M - \mu\mathbf{A}(k)\right)^T\left(\mathbf{I}_M - \mu\mathbf{A}(k)\right)\right\}\tilde{\mathbf{w}}(k) + \\ &2\beta(k)\tilde{\mathbf{w}}^T(k)E\left\{\left(\mathbf{I}_M - \mu\mathbf{A}(k)\right)^T \mathbf{P}(k)\right\}f\left(\mathbf{w}(k)\right) + \\ &\beta^2(k)f^T\left(\mathbf{w}(k)\right)E\left\{\mathbf{P}^T(k)\mathbf{P}(k)\right\}f\left(\mathbf{w}(k)\right).\end{aligned} \tag{63}$$

Setting the derivative of (63) with respect to $\beta(k)$ to zero, we obtain:

$$\beta(k) = -\frac{\tilde{\mathbf{w}}^T(k)E\left\{\left(\mathbf{I}_M - \mu\mathbf{A}(k)\right)^T \mathbf{P}(k)\right\}f\left(\mathbf{w}(k)\right)}{f^T\left(\mathbf{w}(k)\right)E\left\{\mathbf{P}^T(k)\mathbf{P}(k)\right\}f\left(\mathbf{w}(k)\right)}. \tag{64}$$

Recalling the assumption 6, we are able to relax (64) as

$$\beta(k) = -\zeta \frac{\tilde{\mathbf{w}}^T(k)f\left(\mathbf{w}(k)\right)}{\|f\left(\mathbf{w}(k)\right)\|_2^2}, \tag{65}$$

where $\zeta = 1$ for the original $l_1$-NSAF and $l_1$-RNSAF algorithms, and $\zeta = 1 - \mu N/M$ for the proposed $l_1$-qNSAF and $l_1$-qRNSAF algorithms. Since $F(\cdot)$ is a real-valued convex function, by the definition of the sub-gradient in [24], [41], we have

$$\begin{aligned}\tilde{\mathbf{w}}^T(k)f(\mathbf{w}(k)) &\square \left(\mathbf{w}^o - \mathbf{w}(k)\right)^T f(\mathbf{w}(k)) \\ &\leq F(\mathbf{w}^o) - F(\mathbf{w}(k)).\end{aligned} \tag{66}$$

The equal sign in the inequality (66) holds when $\mathbf{w}(k) = \mathbf{w}^o$. Then, inserting (66) into (65) yields an adaptation scheme of $\beta(k)$:

$$\beta(k) = \zeta \frac{\max\left\{\left(F(\mathbf{w}(k)) - F(\mathbf{w}^o)\right), \delta_{\min}\right\}}{\|f(\mathbf{w}(k))\|_2^2}, \tag{67}$$

where $\delta_{\min}$ ($\delta_{\min} > 0$) is small to prevent $\beta(k)$ from freezing. Owing to the unknown vector $\mathbf{w}^o$ in (67), we consider the available estimate $F(\hat{\mathbf{w}})$ instead of $F(\mathbf{w}^o)$, i.e.,

$$\beta(k) = \zeta \frac{\max\left\{\left(F(\mathbf{w}(k)) - F(\hat{\mathbf{w}})\right), \delta_{\min}\right\}}{\|f(\mathbf{w}(k))\|_2^2}. \tag{68}$$

In general, $\mathbf{w}(k)$ is initialized as a null vector, thus to avoid the division in (68) by zero the initial value of $\beta(k)$ is 0.

The current estimate $\mathbf{w}(k)$ among the available estimates is closest to $\mathbf{w}^o$. If we select $\hat{\mathbf{w}} = \mathbf{w}(k)$, then $\beta(k)$ from (68) is always small value $\delta_{\min}$, which is only desirable in the steady-state. Therefore, to get a large $\beta(k)$ to speed up convergence during the transient period, a simple scheme is to update $\hat{\mathbf{w}}$ by $\mathbf{w}(k)$ for every $\lfloor M/N \rfloor$ iterations:

$$\begin{aligned}&\text{if } k = a\lfloor M/N \rfloor,\ a = 0,1,2,...\\ &\quad \hat{\mathbf{w}} \leftarrow \mathbf{w}(k)\\ &else\\ &\quad \hat{\mathbf{w}} \leftarrow 0.5\hat{\mathbf{w}} + 0.5\mathbf{w}(k)\\ &\text{end},\end{aligned} \tag{69}$$

where $\lfloor \cdot \rfloor$ is a floor operator that takes the smallest integer close to its entry. Moreover, $\beta(k)$ given by (68) and (69) remains a small constant in the steady-state, since $\hat{\mathbf{w}} = \mathbf{w}(k) = \mathbf{w}(\infty)$. In practical scenarios, if the prior knowledge of $\mathbf{w}^o$ is also available, then $F(\hat{\mathbf{w}})$ can also be initialized with some constants.

5. Simulation results

Extensive simulations are presented under the fact that the adaptive filter has the same length as the unknown vector. The colored input $u(n)$ is either generated by filtering a white Gaussian signal through a first-order autoregressive system with a pole at 0.9 [23], [36] or a true speech signal. The white Gaussian noise $\eta(n)$ is added to the desired signal, to give a certain signal-to-noise rate (SNR) defined as

$$\text{SNR} = 10\log_{10}\left(E\{y^2(n)\}/\sigma_\eta^2\right),\ y(n) = \mathbf{u}^T(n)\mathbf{w}_o. \tag{70}$$

The cosine modulated filter bank is used and the number of subbands is $N = 4$ or 8 [2]. The expectations concerning the input in evaluating the theoretical results are calculated by ensemble averaging. The quantity, $10\log_{10}[\text{MSD}(k)]$, in (dB), is used to plot these results. All simulated results are the average of 200 independent runs, except for the speech input.

5.1 Comparison of quasi algorithms

In this subsection, we examine the performance of the proposed algorithms by considering two different sparse scenarios.

Example 1: The vector $\mathbf{w}^o$ has $M = 32$ taps; its nonzero coefficients are Gaussian variables with zero mean and unit variance and their positions are randomly selected, with $\|\mathbf{w}^o\|_2 = 1$ [19].

Example 2: The vector $\mathbf{w}^o$ is an acoustic echo channel from Fig. 8(a) in [3], with $M = 512$ taps, also shown in Fig. 2. The channel estimation is crucial for echo cancellation applications.

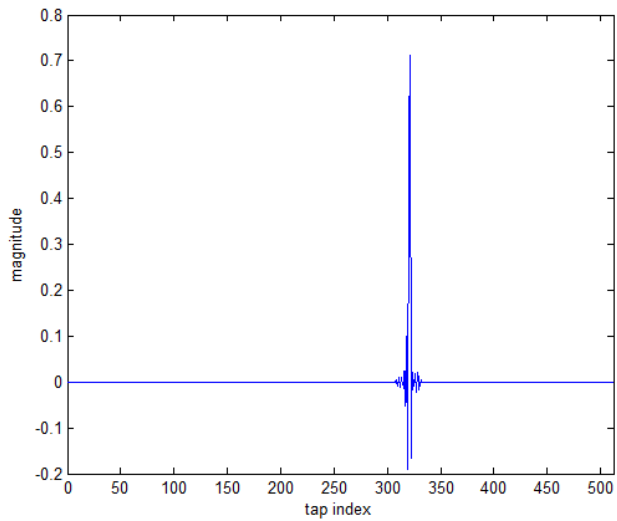

Fig. 2 Acoustic echo channel.

For the NSAF, $l_1$-NSAF, $l_1$-RNSAF, zero-attraction NLMS (ZA-NLMS), reweighted ZA-NLMS (RZA-NLMS), $l_1$-qNSAF and $l_1$-qRNSAF algorithms, Figs. 3 and 4 compare their performance in Examples 1 and 2, respectively. The ZA-NLMS and RZA-NLMS algorithms were presented in [41], which are fullband counterparts of the $l_1$-qNSAF and $l_1$-qRNSAF algorithms, respectively. The parameters of these algorithms are chosen based on the principle of the same convergence rate or steady-state MSD, given in Table 4. As one can see, compared to the NSAF algorithm, these sparsity-aware NSAF algorithms exhibit better performance in the steady-state MSD, since they exploit the sparsity information of the underlying system in the adaptation update. Under the same parameter values, the proposed $l_1$-qNSAF and $l_1$-qRNSAF algorithms have almost the same performance as the $l_1$-NSAF and $l_1$-RNSAF algorithms, respectively. This means that the simplification of (5) and replacement with the expression in (9) is effective in practice. In addition, compared with the fullband ZA-NLMS and RZA-NLMS algorithms, respectively, the proposed $l_1$-qNSAF and $l_1$-qRNSAF algorithms improve the convergence rate, due to the inherent decorrelating property of subband algorithms for correlated input signals. Therefore, considering the limitations of pages, we do not show the results for the existing $l_1$-NSAF and $l_1$-RNSAF algorithms in the following simulations.

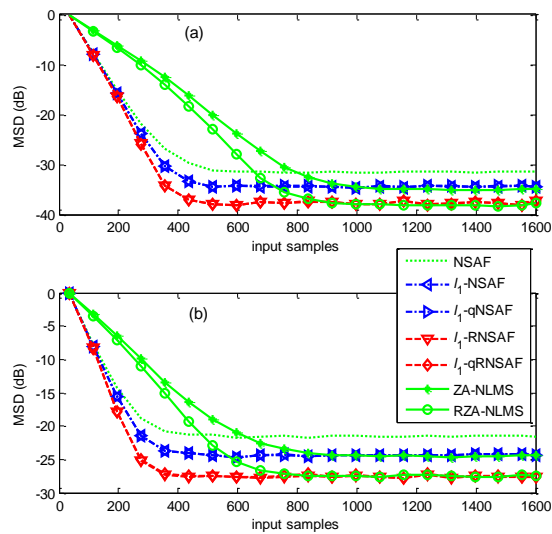

Fig. 3 MSD curves of various algorithms for Example 1 with $Q=2$. (a) SNR=30dB, (b) SNR=20dB. [$N$=4].

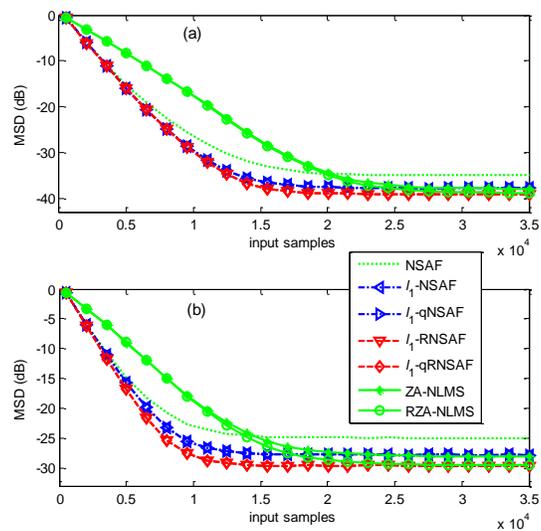

Fig. 4 MSD curves of various algorithms for Example 2. (a) SNR=30dB, (b) SNR=20dB. [$N$=4].

Table 4 Parameters setting of the algorithms.

| Algorithms | Fig. 3 | | Fig. 4 | |
|---|---|---|---|---|
| | SNR=30dB | SNR=20dB | SNR=30dB | SNR=20dB |
| NSAF | $\mu=0.5$ | | $\mu=0.3$ | |
| $l_1$-NSAF | $\mu=0.5$ | | $\mu=0.3$ | |
| | $\beta=4\times10^{-4}$ | $\beta=1\times10^{-3}$ | $\beta=4\times10^{-6}$ | $\beta=5\times10^{-6}$ |
| $l_1$-RNSAF | $\varepsilon=0.05$, $\mu=0.5$ | | $\varepsilon=0.05$, $\mu=0.3$ | |
| | $\beta=5\times10^{-5}$ | $\beta=2\times10^{-4}$ | $\beta=2\times10^{-7}$ | $\beta=7\times10^{-7}$ |
| $l_1$-qNSAF | $\mu=0.5$ | | $\mu=0.3$ | |
| | $\beta=4\times10^{-4}$ | $\beta=1\times10^{-3}$ | $\beta=4\times10^{-6}$ | $\beta=5\times10^{-6}$ |

| | $\varepsilon = 0.05$, $\mu = 0.5$ | | $\varepsilon = 0.05$, $\mu = 0.3$ | |
|---|---|---|---|---|
| $l_1$-qRNSAF | $\beta = 5\times 10^{-5}$ | $\beta = 2\times 10^{-4}$ | $\beta = 2\times 10^{-7}$ | $\beta = 7\times 10^{-7}$ |
| ZA-NLMS | $\mu = 1$ | | | |
| | $\rho = 1\times 10^{-4}$ | $\rho = 3.5\times 10^{-4}$ | $\rho = 5\times 10^{-7}$ | $\rho = 1.8\times 10^{-6}$ |
| RZA-NLMS | $\delta = 0.05$, $\mu = 1$ | | | |
| | $\rho = 2.5\times 10^{-5}$ | $\rho = 9.5\times 10^{-5}$ | $\rho = 3\times 10^{-8}$ | $\rho = 1\times 10^{-7}$ |

5.2 Verification of analyses

Here, we evaluate the effectiveness of the theoretical analyses for the $l_1$-qNSAF and $l_1$-qRNSAF algorithms. The unknown vector is given in Example 1. The parameter $\varepsilon$ in the $l_1$-qNSAF algorithm is set to 0.05, unless otherwise specified.

1) Transient results: the theoretical values for the $l_1$-qNSAF and $l_1$-qRNSAF algorithms are based on the formulas (18), (26) and (28).

Fig. 5 shows the MSD performance of the proposed $l_1$-qNSAF and $l_1$-qRNSAF algorithms using different $\beta$ values. The step size is $\mu = 0.5$. It is clear that the theoretical results match well with those of the simulated results. From Fig. 5, we also find that both proposed algorithms obtain smaller steady-state MSD than the NSAF algorithm as $\beta$ increases; however, when $\beta$ is large than a certain value (e.g., in this case, $\beta = 4\times 10^{-4}$ for the $l_1$-qNSAF and $\beta = 5\times 10^{-5}$ for the $l_1$-qRNSAF), this superiority will be reduced. In other words, there is a particular $\beta$ region, which makes the proposed algorithms outperform the NSAF algorithm in the steady-state performance. Note that, this phenomenon is common for the sparsity-aware algorithms.

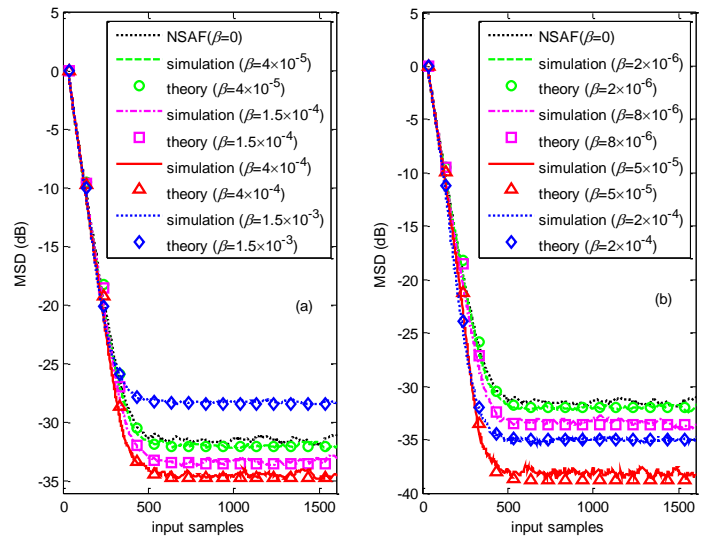

Fig. 5 MSD curves of the proposed algorithms versus $\beta$ values. (a) $l_1$-qNSAF, and (b) $l_1$-qRNSAF. [ $N = 4$, $Q = 2$ and SNR = 30 dB ].

Figs. 6 and 7 depict the MSD performance curves of the proposed algorithms using different step sizes ($\mu = 0.1$ and 0.5) for the cases of SNR=30 dB and SNR=20 dB, respectively. As one can see, the theoretical results are very close to the simulated results.

Moreover, both the proposed algorithms have also a tradeoff problem between fast convergence rate and low steady-state MSD, which is controlled by the fixed step size.

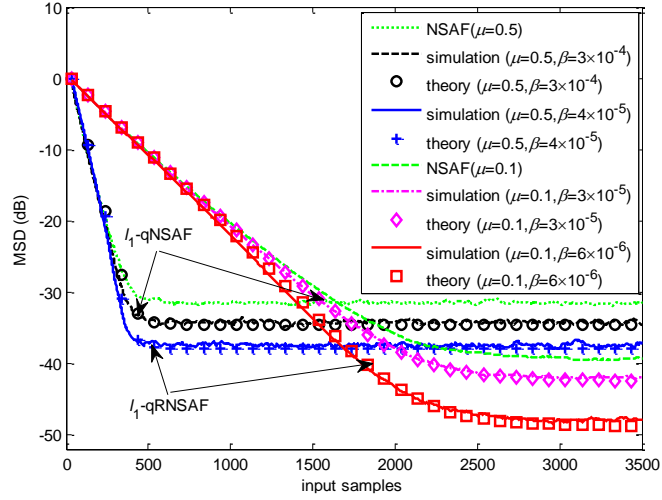

(a)

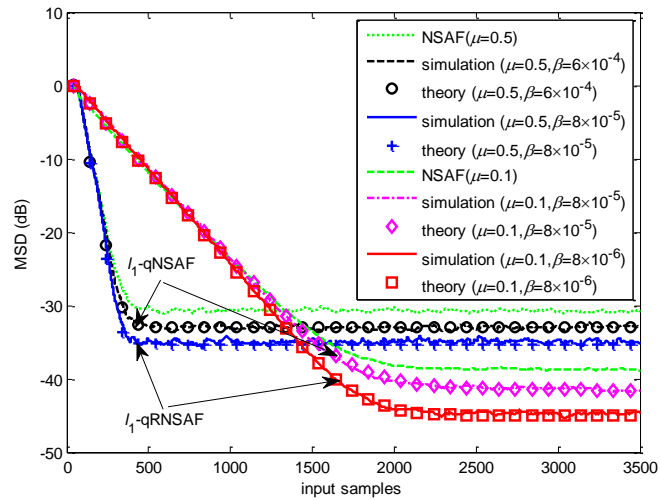

(b)

Fig. 6 MSD curves of the proposed algorithms for different step sizes in the case of SNR=30 dB. (a) $N = 4$ and (b) $N = 8$. [$Q = 2$].

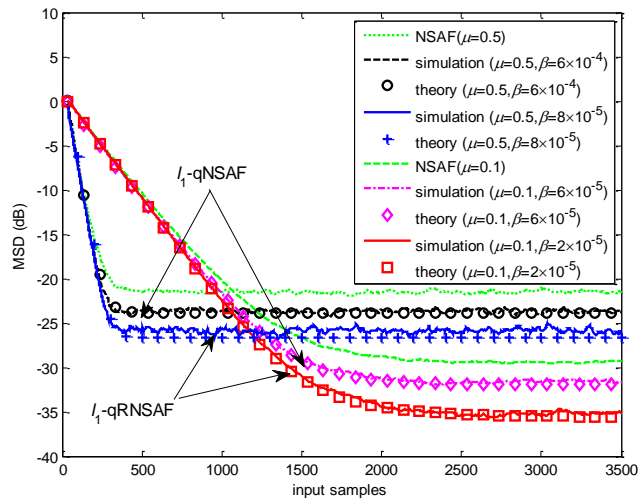

(a)

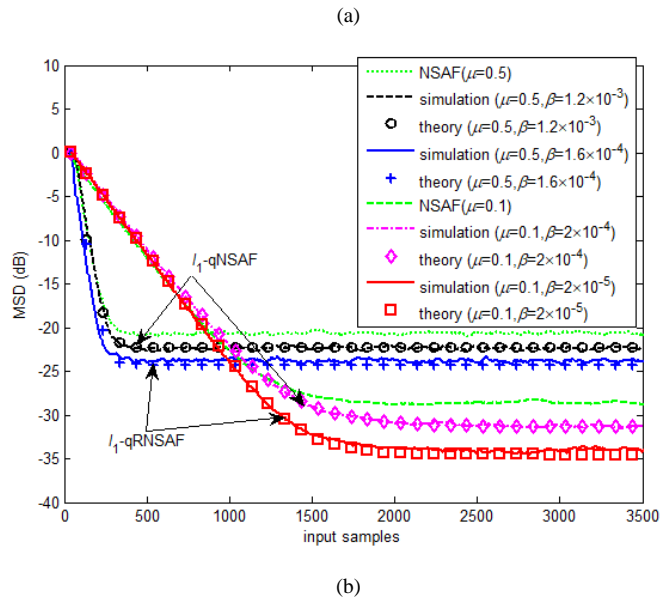

(b)

Fig. 7 MSD curves of the proposed algorithms for different step sizes in the case of SNR $= 20$ dB. (a) $N = 4$ and (b) $N = 8$. [$Q = 2$].

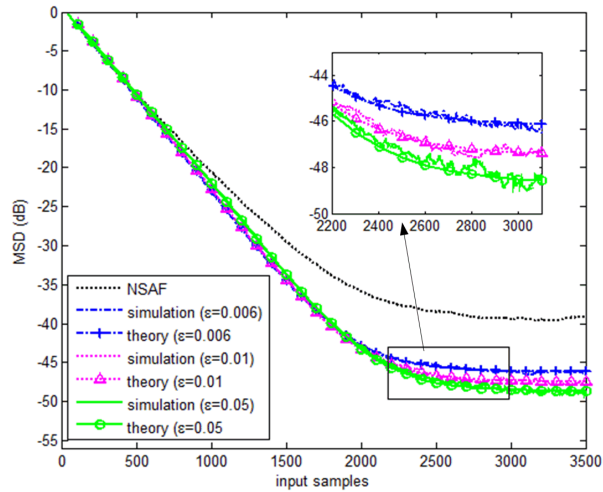

Fig. 8 MSD curves of the $l_1$-qRNSAF algorithm versus different $\varepsilon$ values. [$N = 4$, SNR $= 30$ dB, and $Q = 2$].

Fig. 8 checks the theoretical results of the $l_1$-qRNSAF algorithm using different $\varepsilon$ values, where $\mu$ and $\beta$ are set to $\mu = 0.1$ and $\beta = 8 \times 10^{-6}$, respectively. In this figure, there is a good fitting between theoretical curves and simulated curves. Also, for the $l_1$-qRNSAF algorithm, a large $\varepsilon$ (e.g., $\varepsilon = 0.05$) leads to a lower steady-state MSD than a small $\varepsilon$ (e.g., $\varepsilon = 0.006$), but slows the convergence rate. It is noticeable that for different $\varepsilon$ values in the range [0.006, 0.05], there is an attractive tradeoff between convergence and steady-state performances.

2) Steady-state results: the simulation results are obtained by averaging 500 instantaneous MSD values in the steady-state; the theoretical results are computed according to (60)-(62). Fig. 9 shows the steady-state results of the proposed algorithms with respect to $Q$ and $\beta$. As can been seen, the proposed algorithms will be superior to the NSAF algorithm in the steady-state MSD only when $b$ locates in a particular region, which verifies the result (49). Also, as $Q$ increases, this superiority region narrows. Fig. 10 plots the steady-state MSDs of the proposed algorithms as a function of the step size. The step size $\mu$ varies from 0.01 to 1.0. We set $b = 2 \times 10^{-5}$ for the $l_1$-qNSAF algorithm and $b = 1 \times 10^{-5}$ for the $l_1$-qRNSAF algorithm. It is observed from Fig. 10 that when the parameter $b$ is fixed, the proposed algorithms will have an optimal step size to yield the lowest steady-state MSD. This is because that the superiority region of $b$ is dependent of the step size. Note that, in Figs. 9 and 10, the values obtained by the theory have good agreement with the simulations, even if the assumption 6 is aiming to large $N$.

Without loss of generality, the theory results in Section 3 are also effective for the $l_1$-NSAF and $l_1$-RNSAF algorithms.

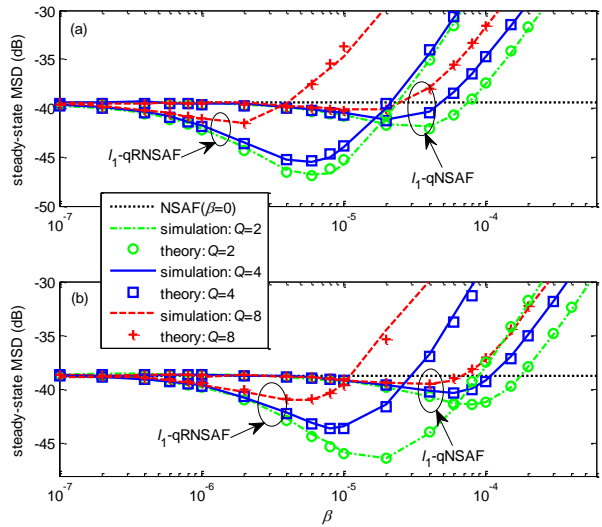

Fig. 9 Steady-state MSDs of the proposed algorithms for different values of $\beta$ and $Q$. (a) $N = 4$ and (b) $N = 8$. [ $\mu = 0.1$, SNR = 30 dB ].

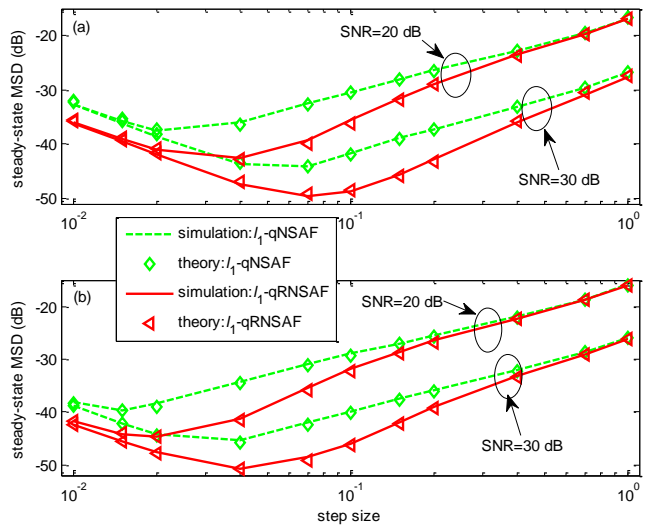

Fig. 10 Steady-state MSDs of the proposed algorithms with different step sizes. (a) $N = 4$ and (b) $N = 8$. [ $Q = 2$ ].

5.3 Adaptation intensity algorithms

In this subsection, we evaluate the proposed $l_1$-qNSAF and $l_1$-qRNSAF algorithms with adaptation of $\beta(k)$ according to (68) and (69), called respectively here the A-$l_1$-qNSAF and A-$l_1$-qRNSAF algorithms, for estimating acoustic echo channel given in Fig. 2.

To compare the tracking capability, an abrupt change of echo channel occurs at the middle of input samples, by shifting the impulse response to the right by 12 taps [7], [45]. The same step size for all the algorithms is $\mu = 0.5$.

Fig. 11 illustrates the effect of $\delta_{\min}$ on the performance of the A-$l_1$-qNSAF and A-$l_1$-qRNSAF algorithms. One can find that $\delta_{\min}$ influences the performance of the algorithms, but its sensitivity is weaker than the effect of $\beta$ on the performance of the $l_1$-qNSAF and $l_1$-qRNSAF algorithms. This is due to the fact that even in a simple case of $\delta_{\min} = 0$, the A-$l_1$-qNSAF and A-$l_1$-qRNSAF algorithms are superior to the NSAF algorithm for sparse system identification.

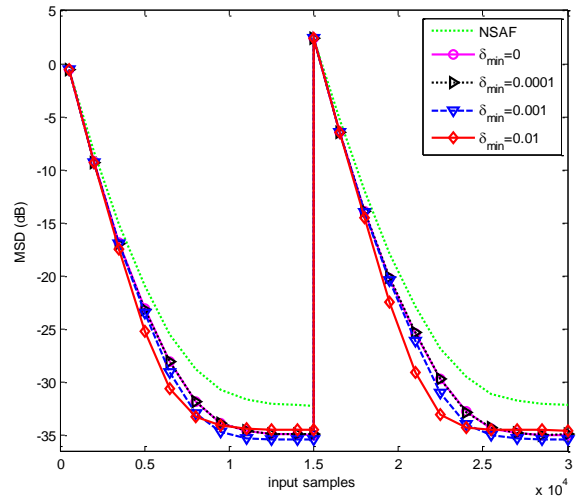

(a)

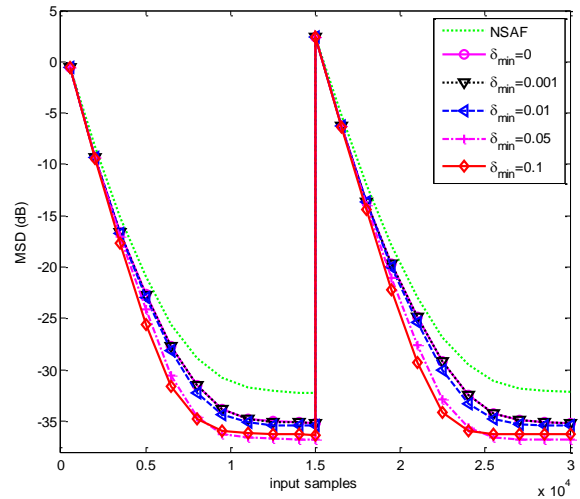

(b)

Fig. 11 MSD curves of the (a) A-$l_1$-qNSAF and (b) A-$l_1$-qRNSAF algorithms using different $\delta_{\min}$ values. [ $N = 4$, SNR = 30 dB ].

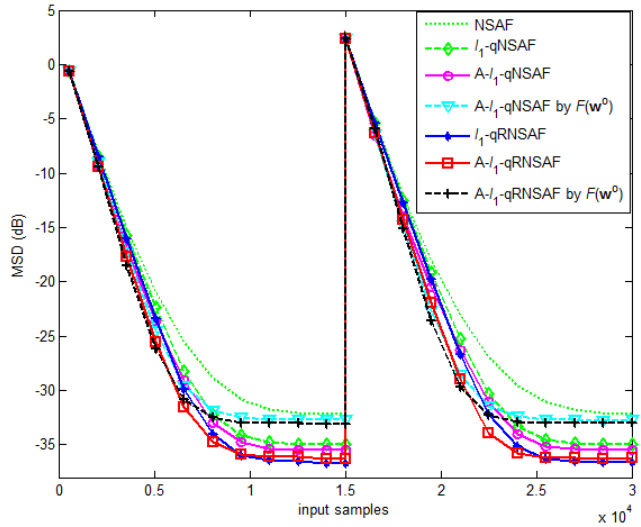

(a)

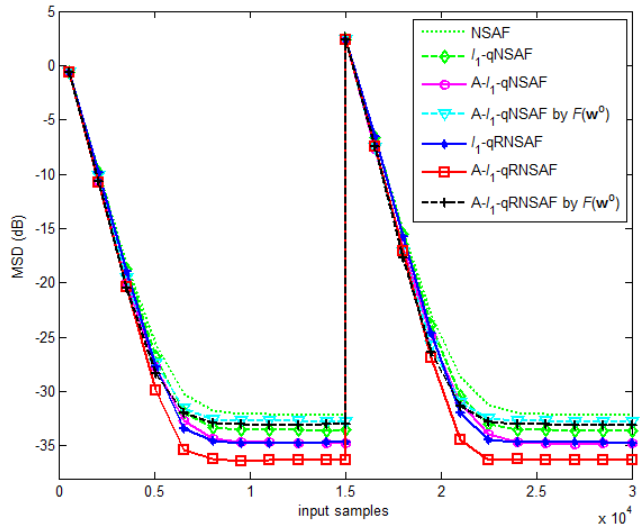

(b)

Fig. 12 MSD curves of various algorithms for acoustic echo channel identification. (a) $N = 4$ and (b) $N = 8$. [ SNR = 30 dB ].

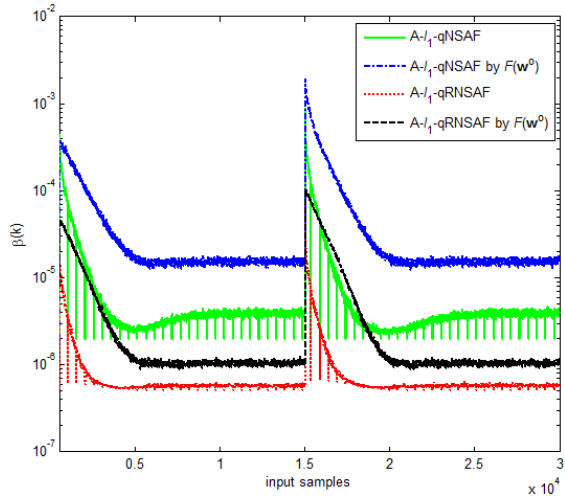

(a)

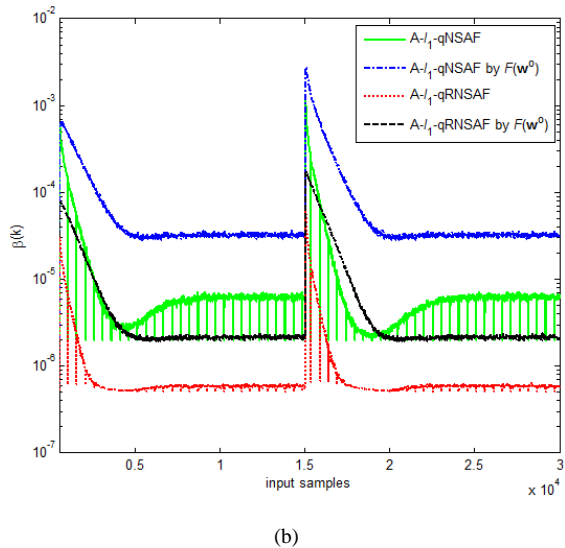

(b)

Fig. 13 Evolution of $\beta(k)$ for the algorithms involved in Fig. 12. (a) $N=4$ and (b) $N=8$.

In Fig. 12, we compare the performance of the A-$l_1$-qNSAF and A-$l_1$-qRNSAF algorithms with that of the $l_1$-qNSAF and $l_1$-qRNSAF algorithms. Here, we set $\beta=3\times10^{-6}$ and $\beta=3\times10^{-7}$ for the $l_1$-qNSAF and $l_1$-qRNSAF algorithms, respectively, to outperform the NSAF algorithm as possible. For the A-$l_1$-qNSAF and A-$l_1$-qRNSAF algorithms, we set $\delta_{\min}=0.001$ and $\delta_{\min}=0.1$, respectively. As can be seen from Fig. 12, the $l_1$-qNSAF and $l_1$-qRNSAF algorithms with the adaptation intensity parameter have better performance than those with the fixed intensity parameter. Intuitively, the ideal adaptation intensity scheme that $\mathbf{w}^o$ in (67) is known, should work better than the estimated approach (69). As a matter of fact, (69) may be better in the steady-state performance. This reason may be twofold: 1) the upper bound in (66) is used to derive the adaptation scheme (67); 2) $F(\mathbf{w}^o)$ is farther away from $F(\mathbf{w}(k))$ than $F(\hat{\mathbf{w}})$ so that the ideal adaptation intensity scheme yields larger $\beta(k)$ values than the approach (69), also shown in Fig. 13. Fig. 13 depicts the evolution of $\beta(k)$ for the A-$l_1$-qNSAF and A-$l_1$-qRNSAF algorithms in Fig. 12. Moreover, in Fig. 13, the curves of $\beta(k)$ (A-$l_1$-qNSAF and A-$l_1$-qRNSAF) have many burrs, because $\hat{\mathbf{w}}$ in (68) is replaced periodically with $\mathbf{w}(k)$ for every $\lfloor M/N \rfloor$ iterations, i.e., (69).

In addition, using a true speech signal as the input, Fig. 14 also illustrates the effectiveness of the proposed adaptation intensity scheme for the sparsity-aware NSAF algorithms. In this example, we add a regularization constant 0.1 in the denominator of the update (9) to avoid the division by zero, and other parameters' setting is the same as Fig. 12. We remark that other related approaches such as those in [46], [47], [48], [49] and [50] could be also further investigated.

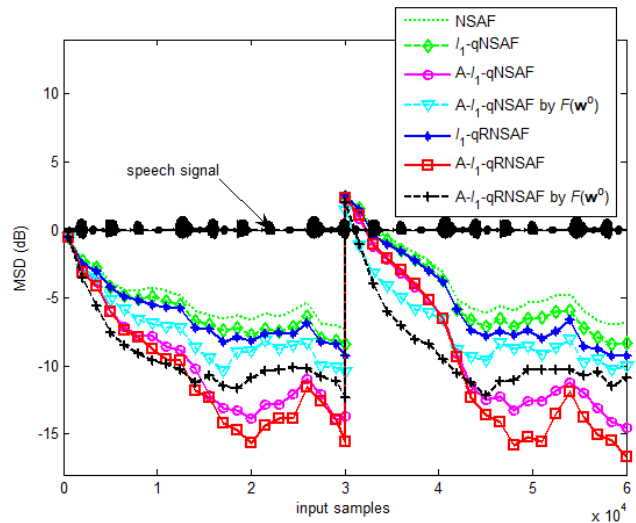

(a)

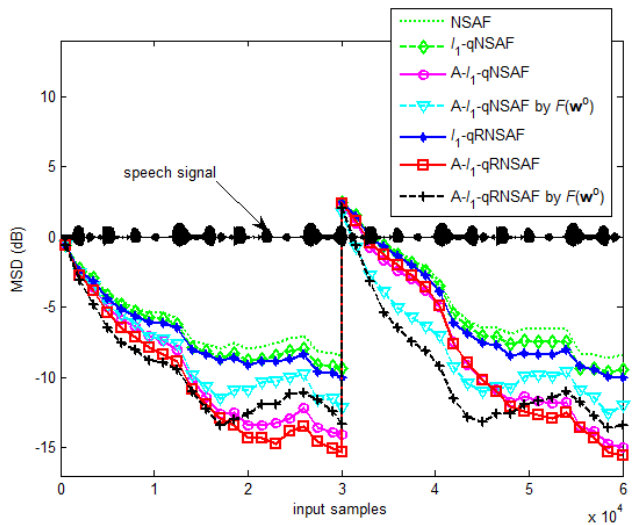

(b)

Fig. 14 MSD curves of various algorithms for identifying acoustic echo channel with speech input (single realization). (a) $N=4$ and (b) $N=8$.

[ SNR = 30 dB ].

6. Conclusions

This paper has developed the $l_1$-qNSAF and $l_1$-qRNSAF algorithms for sparse system identification by using the gradient descent method or relaxing the projection operation of $f(\mathbf{w}(k))$ in (5). The proposed algorithms reduce the computational cost of the $l_1$-NSAF and $l_1$-RNSAF algorithms presented in [23], respectively, while retaining the same convergence performance. We present a detailed performance analysis for these four sparsity-aware NSAF algorithms. Due to the use of the vectorization argument and Kronecker product in the analysis, we have derived expressions for the transient and steady-state MSD without restricting the distribution of the input signal to be a particular model. Moreover, to further improve the performance of the proposed algorithms, an adaptive method for choosing the intensity parameter is also designed. Simulation results in various environments have demonstrated our algorithms and theoretical analysis.

Appendix A: Vectorization of I-IV terms in (20)

Term I:

$$\begin{aligned}
&\text{vec}\left(E\left\{(\mathbf{I}_M - \mu\mathbf{A}(k))\mathbf{\Phi}(k)(\mathbf{I}_M - \mu\mathbf{A}(k))^T\right\}\right) \\
&= E\left\{\text{vec}\left((\mathbf{I}_M - \mu\mathbf{A}(k))\mathbf{\Phi}(k)(\mathbf{I}_M - \mu\mathbf{A}(k))^T\right)\right\} \\
&= E\left\{\left[(\mathbf{I}_M - \mu\mathbf{A}(k)) \otimes (\mathbf{I}_M - \mu\mathbf{A}(k))\right]\text{vec}(\mathbf{\Phi}(k))\right\} \\
&= E\left\{(\mathbf{I}_M - \mu\mathbf{A}(k)) \otimes (\mathbf{I}_M - \mu\mathbf{A}(k))\right\}\text{vec}(\mathbf{\Phi}(k)),
\end{aligned} \quad (A1)$$

Term II:

$$\begin{aligned}
&\text{vec}\left(E\left\{(\mathbf{I}_M - \mu\mathbf{A}(k))\mathbf{H}(k)\mathbf{P}^T(k)\right\}\right) \\
&= E\left\{\text{vec}\left((\mathbf{I}_M - \mu\mathbf{A}(k))\mathbf{H}(k)\mathbf{P}^T(k)\right)\right\} \\
&= E\left\{\left[\mathbf{P}(k) \otimes (\mathbf{I}_M - \mu\mathbf{A}(k))\right]\text{vec}(\mathbf{H}(k))\right\} \\
&= E\left\{\mathbf{P}(k) \otimes (\mathbf{I}_M - \mu\mathbf{A}(k))\right\}\text{vec}(\mathbf{H}(k)),
\end{aligned} \quad (A2)$$

Term III:

$$\begin{aligned}
&\text{vec}\left(E\left\{\mathbf{P}(k)\mathbf{H}(k)(\mathbf{I}_M - \mu\mathbf{A}(k))^T\right\}\right) \\
&= E\left\{\text{vec}\left(\mathbf{P}(k)\mathbf{H}^T(k)(\mathbf{I}_M - \mu\mathbf{A}(k))^T\right)\right\} \\
&= E\left\{\left[(\mathbf{I}_M - \mu\mathbf{A}(k)) \otimes \mathbf{P}(k)\right]\text{vec}(\mathbf{H}^T(k))\right\} \\
&= E\left\{(\mathbf{I}_M - \mu\mathbf{A}(k)) \otimes \mathbf{P}(k)\right\}\text{vec}(\mathbf{H}^T(k)),
\end{aligned} \quad (A3)$$

Term IV:

$$\begin{aligned}
&\text{vec}\left(E\left\{\mathbf{P}(k)\mathbf{\Theta}(k)\mathbf{P}^T(k)\right\}\right) \\
&= E\left\{\text{vec}\left(\mathbf{P}(k)\mathbf{\Theta}(k)\mathbf{P}^T(k)\right)\right\} \\
&= E\left\{\left[\mathbf{P}(k) \otimes \mathbf{P}(k)\right]\text{vec}(\mathbf{\Theta}(k))\right\} \\
&= E\left\{\mathbf{P}(k) \otimes \mathbf{P}(k)\right\}\text{vec}(\mathbf{\Theta}(k)).
\end{aligned} \quad (A4)$$

Appendix B: Verification of assumption 4

Fig. 15 shows the simulated probability density function (pdf) of the weight error components $\tilde{w}_m(k)$ in the $l_1$-qNSAF algorithm (with $\beta = 4\times 10^{-4}$), for different iterations $k = 80$ (transient-state) and $k = 500$ (steady-state), and different coefficients $m = 5$ (large coefficient) and $m = 10$ (zero coefficient). As one can see, for large and zero coefficients, the corresponding weight error components have an approximately Gaussian pdf. Similar results can also be observed in the $l_1$-qRNSAF algorithm. Therefore, the assumption 4 is reasonable.

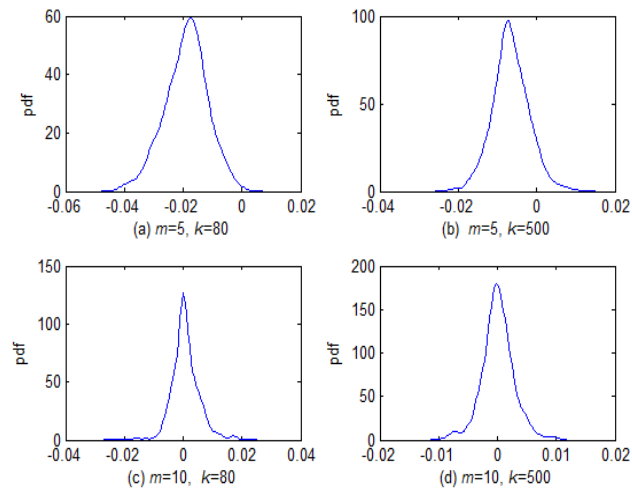

Fig. 15 Simulated pdfs of the weight error components $\tilde{w}_m(k)$, with 500 independent runs. Other simulation setting is the same as Fig. 5.

Acknowledgement

This work was partially supported by National Nature Science Foundation of P.R. China (Grant: 61871461, 61571374, and 61433011).

**Yi Yu** was born in Sichuan Province, China, in 1989. He received the B.E. degree at XiHua University, Chengdu, China, in 2011, and the M.S. degree and Ph.D. degree at Southwest Jiaotong University, Chengdu, China, in 2014 and 2018, respectively. From Dec. 2016 to Dec. 2017, he was a visiting Ph.D. student with the Department of Electronic Engineering, University of York, United Kingdom. He is currently an Associate Professor with the School of Information Engineering,

Southwest University of Science and Technology, Mianyang, China. His research interests include adaptive signal processing, distributed estimation, and compressive sensing.

**Haiquan Zhao** was born in Henan Province, China, in 1974. He received the B.S. degree in applied mathematics, and the M.S. and Ph.D degrees in signal and information processing from Southwest Jiaotong University, Chengdu, China, in 1998, 2005, and 2011, respectively. Since 2012, he was a Professor with the School of Electrical Engineering, Southwest Jiaotong University. His current research interests include adaptive filtering algorithm, nonlinear active noise control and nonlinear system identification. He has authored or co-authored over 100 journal papers and holds 13 invention patents.

**Rodrigo C. de Lamare** was born in Rio de Janeiro, Brazil, in 1975. He received his Diploma in electronic engineering from UFRJ in 1998 and the MSc and PhD degrees in electrical engineering from the Pontifical Catholic University of Rio de Janeiro (PUC-RIO) in 2001 and 2004, respectively. He then worked as a Postdoctoral Fellow from January to June 2005 at the Centre for Telecommunications Studies (CETUC), PUC-RIO and from July 2005 to January 2006 at the Signal Processing Laboratory, UFRJ. Since January 2006, he has been with the Communications Group, Department of Electronics, University of York, United Kingdom, where he is a Professor. Since April 2013, he has also been a Professor at PUC-RIO. Dr de Lamare has participated in numerous projects funded by government agencies and industrial companies. He received a number of awards for his research work and has served as the general chair of the IEEE 7th International Symposium on Wireless Communication Systems (ISWCS) 2010, held in York, UK in September 2010, as the technical programme chair of ISWCS 2013 and WSA 2015 in Ilmenau, Germany, and as the general chair of the 9th IEEE Sensor Array and Multichannel Signal Processing Workshop (SAM) in Rio de Janeiro, Brazil, in July 2016. Dr de Lamare is a senior member of the IEEE and an elected member of the IEEE Signal Processing Theory and Method technical committee. He currently serves as editor for IEEE Transactions on Communications, IEEE Wireless Communications Letters, and as a senior area editor for the IEEE Signal Processing Letters. His research interests lie in communications and signal processing, areas in which he has published nearly 400 papers in international journals and conferences.

**Lu Lu** was born in Chengdu, China, in 1990. He received the Ph.D. degree in the field of signal and information processing at the School of Electrical Engineering, Southwest Jiaotong University, Chengdu, China, in 2018. From 2017 to 2018, he was a visiting Ph.D. student with the Electrical and Computer Engineering at McGill University, Montreal, QC, Canada. He is currently a Postdoctoral Fellow with the College of Electronics and Information Engineering, Sichuan University, Chengdu, China. His research interests include adaptive filtering, kernel methods and distributed estimation.